\documentclass[%
reprint,
amsmath,amssymb,
aps,
]{revtex4-1}
\usepackage{physics}
\usepackage[caption=false]{subfig}
\usepackage{graphicx}
\usepackage{dcolumn}
\usepackage{bm}
\usepackage{xcolor}
\usepackage{float}
\newcommand{\avg}[1]{\left<#1\right>}
\newcommand{\h}[1]{\mathcal{H}_{#1}}
\begin{document}
	
	\preprint{}

	\title{Theory of nonlinear interactions between x rays and optical radiation in crystals}
	
	
	\author{R. Cohen and S. Shwartz}
	\affiliation{Physics Department and Institute of Nanotechnology, Bar-Ilan University, Ramat Gan 52900, Israel}
	
	
	\begin{abstract}
		
		We show that the nonlinear interactions between x rays and longer wavelengths in crystals depend strongly on the band structure and related properties. Consequently, these types of interactions can be used as a powerful probe for fundamental properties of crystalline bulk materials. In contrast to previous work that highlighted that these types of nonlinear interactions can provide microscopic information on the valence electrons at the atomic scale resolution, we show that these interactions also contain information that is related to the periodic potential of the crystal. We explain how it is possible to distinguish between the two contributions. Our work indicates on the possibility for the development of novel multi-dimensional pump-probe metrology techniques that will provide spectroscopic information combined with structural information including  ultrafast dynamics at the atomic scale.      
		
	\end{abstract}
	
	
	\maketitle
	
	%
	%
	%
	
	\section{\label{sec:level1}Introduction}
	
	The possibility to utilize nonlinear interactions between x rays and radiation at wavelengths ranging from infrared (IR) to ultraviolet (UV) as an atomic-scale probe for inter-molecular interactions and for properties of valence electrons has been discussed in several publications \cite{danino1981parametric, tamasaku2009determining, tamasaku2011visualizing, glover2012x, barbiellini2015explaining, schori2017parametric, rouxel2018x, borodin2019evidence, sofer2019observation}. This probe can provide a new insight into atomic-scale processes, such as charge transfer between atoms in molecules and microscopic redistribution of charges in response to illumination. The technique relies on the atomic scale wavelengths of the x-rays, which provide the high resolution, while the longer wavelengths (UV/optical) are used to enhance the interactions with the valence electrons, which are usually weak for x rays. 
	
	The strong wavelength dependence of spontaneous parametric down conversion (SPDC) of x rays into UV, which has been observed in recent experiments \cite{tamasaku2007idler,tamasaku2011visualizing,borodin2017high,schori2017parametric, borodin2019evidence, sofer2019observation} suggests that nonlinear interactions between x rays and longer wavelengths can be used also as new spectroscopy tools for the investigation of phenomena that traditionally are probed by using long wavelength radiation with the advantage of providing microscopic atomic-scale information. Moreover, x-rays penetrate into materials more than electrons and therefore provide bulk properties in contrast to methods that are relied on electron or ion scattering.
	
	The main challenge in measuring x ray and optical/UV nonlinear interactions is the weakness of the effects. To overcome this challenge, experiments are performed with crystals where the periodic structure is used to enhance the signal in analogy to Bragg scattering, in which the reciprocal lattice vector is used to achieve momentum conservation (phase matching). The use of the reciprocal lattice vector also provides the atomic scale resolution, which can be achieved by measuring the efficiencies of the effect for many reciprocal lattice vectors and by using Fourier analysis \cite{tamasaku2011visualizing}. Since the goal of the measurements is to probe microscopic information, the use of crystals introduces a new challenge for the interpretation of the results since the measured signal depends not just on the atomic or inter unit cell interactions between the valence electrons but also on the periodic potential of the materials. It is therefore essential to develop a formalism that enables the separation of the two contributions.

	We note that other approaches such as inelastic x-ray scattering (IXS) are very useful for the investigation of properties of valence electrons including for instance valence charge distribution, density of states, and the macroscopic linear response \cite{sinha2001theory,wang2012lindhard,baron2015introduction}. In these types of processes, core electrons are excited to valance states and decay by non-radiative scattering processes to lower valence states. In the last step of the process, the electrons return to the original core state and radiate at lower photon energy with an energy that corresponds to the energy difference between the two intermediate valence states.     
	However, despite the similarly in the measurements to nonlinear x-ray interactions (in both x-ray photons at lower energies are generated and measured), IXS processes are linear and require the excitation of core electrons, while nonlinear processes are parametric (the energy transfer between the electromagnetic field and the electronic system during the entire process is negligible). As we will show below, the nonlinear current density includes only matrix elements that do not mix between core and valence states in contrast to IXS. More important, the use of the reciprocal lattice vector for phase matching in the nonlinear processes provides the possibility to utilize Fourier synthesis to measure atomic scale structural information in contrast to conventional IXS.

	To date, most theoretical models that have been considered for the description of nonlinear interactions between x rays and longer wavelengths have focused on the ability to observe microscopic information and on the estimation of the strength of the effects
	\cite{freund1970optically,freund1971resonant,freund1972nonlinear,tamasaku2009determining,tamasaku2011visualizing,glover2012x,dorfman2012photon,barbiellini2015explaining,rouxel2018x}. However, they have not addressed the challenge of the separation of the inter unit cell information from the periodic information. We note that Freund and Levine and also Jha and colleagues \cite{freund1968nonlinear,jha1968nonlinear,freund1969parametric,freund1970optically,freund1971resonant,freund1972nonlinear,jha1972nonlinear1,jha1972nonlinear2}
	have considered the periodic structure but not the influence of the electronic band structure on the nonlinear interactions, 
	which can be significant for valence electrons with binding energies that are weaker or on the order of the periodic potential. In a recent theoretical paper, the authors analyzed the general case of x ray diffraction from laser driven systems using a quantum electrodynamics approach, but in that paper the main focus is on the nonperturbative regime \cite{popova2018theory}. 
	
	In several recent experimental papers, the experimental results were fitted to the theory by using the band gap rather than the atomic binding energy, but although good agreements were obtained for transparent materials, this approach is phenomenological and failed for energies near or above the band gap \cite{schori2017parametric}.
	
	In this paper, we illustrate how the nonlinear interaction between x-rays and longer wavelengths depends on the band structure and related properties by using perturbation theory in the density matrix formalism. 
	As an example we describe explicitly the dependence of the nonlinear interactions on the joint density of states of interband transitions.
	In addition, we analyze the polarization dependencies of the nonlinear interactions and predict that it is not trivial.
	
	\section{\label{sec:level2}Theoretical description and general formalism}
\begin{figure}[ht]
		\subfloat[
		\label{sfig:SFG}]{%
			\includegraphics[scale=0.4]{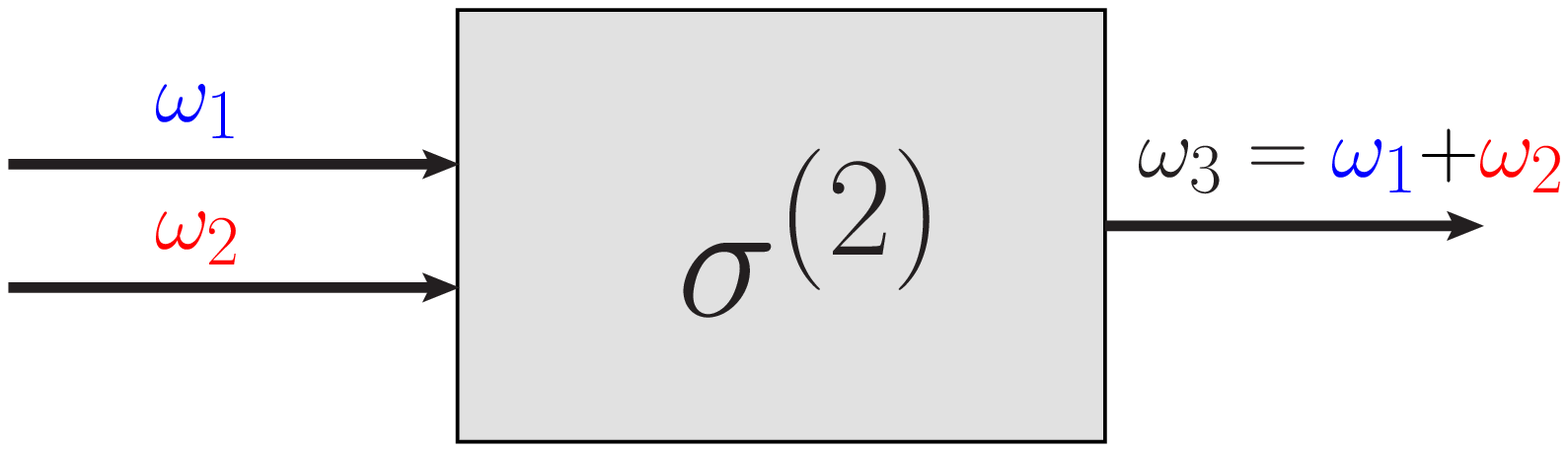}%
		}\hfill
		\subfloat[
		\label{sfig:SPDC}]{%
			\includegraphics[scale=0.4]{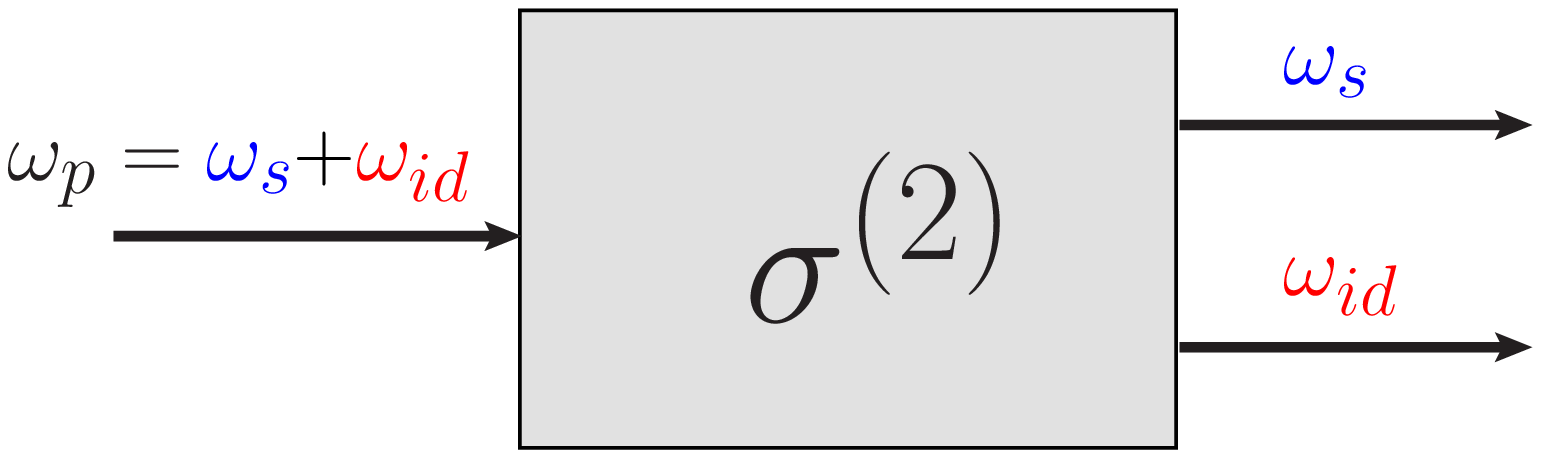}%
		}
		\caption{Schematic diagrams for the nonlinear processes of SFG and SPDC.
			(a) SFG: Two input photons at frequencies $\omega_1$ and $\omega_2$ are converted to a photon at frequency $\omega_3=\omega_1+\omega_2$.
			(b) SPDC: An input pump photon at frequency $\omega_p$ is converted to a photon pair at frequencies $\omega_s$ and $\omega_{id}$.
		}
		\label{fig:SFG-SPDC}
	\end{figure}

	We focus here on the second order nonlinearity that constitutes a source term in the wave equations that describe effects such as sum-frequency generation (SFG) and SPDC.
	Schematic diagrams for these processes are presented in Fig. \ref{fig:SFG-SPDC}. 
	The nonlinear interactions can be 
	described by the second nonlinear current density or the nonlinear conductivity.
	\subsection{Hamiltonian}
	To describe the nonlinear optical interaction in the crystal, we use 
	the minimal electromagnetic coupling Hamiltonian of an electron in a periodic potential
	\begin{eqnarray}
	\begin{aligned}
	&
	\h{}=\h{0}+\h{\text{int}}
	=\frac{(\vb{p}+e\vb{A})^2}{2m}+V(\vb*{x}),\\&
	\h{0}=\frac{\vb{p}^2}{2m}+V(\vb*{x}),
	\,
	\h{\text{int}}=\frac{e}{2m}\left(\vb{p}\cdot\vb{A}+\vb{A}\cdot\vb{p}\right)+\frac{e^2}{2m}\vb{A}^2,
	\end{aligned}
	\end{eqnarray}
	where the unperturbed Hamiltonian is $\h{0}$, 
	the electron mass is $m$, the momentum operator is $\vb{p}$
	and the potential $V(\vb*{x})=V(\vb*{x}+\vb{R})$
	is periodic with respect to translation by a lattice vector, where $\vb{R}$ is any lattice vector.
	In our derivation, the electromagnetic interaction described by $\h{\text{int}}$ is treated as a perturbation, 
	where 
	$-e$ is the electron charge,
	$\vb{A}(\vb*{x},t)=\sum\limits_{l}\frac{\vb*{\varepsilon}(\omega_l)}{i\omega_l} e^{-i\omega_l t}$ is the electromagnetic vector potential assumed in the Coulomb gauge, and $\vb*{\varepsilon}(\omega_l)$ is the electric field envelope of the l-th mode.
	The solutions of the unperturbed Hamiltonian $\h{0}$ are given by the Bloch states
	\begin{equation}
	\h{0}\ket{n\,\vb{q}}=\varepsilon_n(\vb{q})\ket{n\,\vb{q}},
	\end{equation}
	with eigenvalues $\varepsilon_{n}(\vb{q})$, where $n$ is the band number and $\vb{q}$ is the wave vector associated with the Bloch state.
	\subsection{Density matrix}
	We follow the standard procedure for the calculation of nonlinear conductivity by perturbation theory in the density matrix formalism
	\cite{jha1968nonlinear}.
	The density matrix elements $\rho_{nm}$ evolve according to
	\begin{eqnarray}
	i\hbar \pdv{\rho_{nm}}{t}
	=\left[\mathcal{H}_0+\mathcal{H}_{\text{int}},\rho\right]_{nm}
	-i\hbar\gamma_{nm}\left(
	\rho_{nm}-\rho^{(0)}_{nm}
	\right)
	\label{rho-dynamics}
	\end{eqnarray}
	where $n$ and $m$ signifies the eigenstates of the unperturbed Hamiltonian $\mathcal{H}_0$.
	We denote the phenomenological damping coefficients as $\gamma_{nm}$
	and $\rho^{(0)}=f_0(\mathcal{H}_0)$ is the relaxed density matrix which is the normalized Fermi-Dirac distribution, such that $\tr(\rho^{(0)})=1$.
	The full density matrix $\rho(t)$, 
	can be expressed as a power series in the field $\vb{A}$
	\begin{eqnarray}
	\rho(t)=\rho^{(0)}+\rho^{(A)}+\rho^{(AA)}+\cdots,
	\label{rho-order}
	\end{eqnarray}
	where the first term on the right hand side is the relaxed density matrix in the absence of the field, the second term is linear with the field amplitude, the third term is square in the field amplitude and so on.
	
	\subsection{Current density}
	Following the perturbation theory formalism, we calculate average current density by using  
	\begin{eqnarray}
	\begin{aligned}
	\avg{\vb*{j}}(\vb*{x},t)=\tr(\rho(t)\vb*{j}_{\text{op}}(\vb*{x},t)),
	\end{aligned}
	\end{eqnarray}
	where the current density operator \cite{bloembergen1964quantum} is given by
	$\vb*{j}_{\text{op}}
	(\vb*{x},t)=
	\vb*{j}_{\text{op}}^{\vb*{p}}
	(\vb*{x})
	+
	\vb*{j}_{\text{op}}^{\vb{A}}
	(\vb*{x},t)
	$.
	The first term being the momentum current density 
	$
	\vb*{j}_{\text{op}}^{\vb*{p}}(\vb*{x})=
	\frac{1}{2m}
	\left\{
	D_{\text{op}}(
	\vb*{x})
	,
	\vb*{p}
	\right\}
	$ 
	and the second term is the so called gauge current density 
	$
	\vb*{j}_{\text{op}}^{\vb{A}}(\vb*{x},t)
	=
	\frac{e}{m}
	D_{\text{op}}(
	\vb*{x})
	\vb{A}(\vb*{x},t)
	$.
	We expressed these operators by the charge density operator, which we define as 
	$
	D_{\text{op}}(
	\vb*{x})
	=
	-e
	\ket{\vb*{x}}
	\bra{\vb*{x}}
	$.
	The current density can be expressed as a power series in the field $\vb{A}$
	\begin{eqnarray}
	\begin{aligned}&
	\avg{\vb*{j}}(\vb*{x},t)=
	\sum\limits_{n}
	\avg{\vb*{j}}^{(n)}
	\\&=
	\avg{\vb*{j}}^{(0)}(\vb*{x})+
	\avg{\vb*{j}}^{(1)}(\vb*{x},t)+
	\avg{\vb*{j}}^{(2)}(\vb*{x},t)+\cdots,
	\end{aligned}
	\end{eqnarray}
	where n signifies the order in which the current density depends on the field.
	Following Eq. (\ref{rho-order}) we find that the second order current density is
	$\avg{\vb*{j}}^{(2)}(\vb*{x},t)=
	\avg{\vb*{j}}^{(2,\vb*{p})}
	+
	\avg{\vb*{j}}^{(2,\vb{A})}
	$
	, where
	$
	\avg{\vb*{j}}^{(2,\vb*{p})}=
	\tr(\rho^{(AA)}\,
	\vb*{j}_{\text{op}}^{\vb*{p}}
	(\vb*{x})
	)
	$
	is the second order current density that arises from the momentum part
	and
	$
	\avg{\vb*{j}}^{(2,\vb{A})}=
	\tr(\rho^{(A)}\,
	\vb*{j}_{\text{op}}^{\vb{A}}
	(\vb*{x},t))
	$
	is the second order current density that arises from the gauge part of the current density operator.
	The temporal part of the second order current density can be Fourier expended
	\begin{eqnarray}
	\avg{\vb*{j}}^{(2)}(\vb*{x},t)
	=\frac{1}{2}\sum\limits_{l,l'}\avg{\vb*{j}}^{(2)}(\vb*{x};\omega_l+\omega_{l'})e^{-i(\omega_l+\omega_{l'}) t}
	.
	\label{current-fourierexpansion}
	\end{eqnarray}
	Each Fourier coefficient in the direction of the generated field, is related to the conductivity
	\cite{callaway2013quantum} by the following expression
	\begin{eqnarray}&
	\begin{aligned}
	&
	\avg{\vb*{j}}^{(2)}(\vb*{x};\omega_l+\omega_{l'})\cdot \hat{e}_i\\&=
	\sum\limits_{j,k=1}^{3}
	\sigma_{ijk}^{(2)}(\vb*{x};\omega_l+\omega_{l'})
	\left(
	\vb*{\varepsilon}(\omega_l)
	\cdot
	\hat{e}_j
	\right)
	\left(
	\vb*{\varepsilon}(\omega_{l'})
	\cdot
	\hat{e}_k
	\right)
	\end{aligned}
	\label{current-conductivity-relation}
	\end{eqnarray}
	where $\hat{e}_i$ is a unit vector in the i-th direction and $\left\{i,j,k\right\}$ indicates Cartesian coordinates.
	Since the unperturbed system is periodic in the lattice spacing, we Fourier expand spatial part of the nonlinear conductivity, 
	\begin{eqnarray}
	\begin{aligned}
	&
	\sigma_{ijk}^{(2)}(\vb*{x};\omega_l+\omega_{l'})=
	\sum\limits_{\vb{G}}
	\sigma_{ijk}^{(2)}(\omega_l+\omega_{l'};\vb{G})
	e^{i\vb{G}\cdot\vb*{x}}\,,
	\\&
	\sigma^{(2)}_{ijk}(\omega_l+\omega_{l'};\vb{G})=
	\frac{1}{V}\int\limits_{V}d\vb*{x}\,
	\sigma_{ijk}^{(2)}(\vb*{x};\omega_l+\omega_{l'})
	e^{-i\vb{G}\cdot\vb*{x}}\,.
	\end{aligned}
	\label{conductivity-spatialfouriercomponent}
	\end{eqnarray}
	where $\vb{G}$ is the reciprocal lattice vector and $V$ is the crystal volume. The spatial Fourier component of the conductivity can be selected via the phase-matching condition relevant to the process of choice.

	\subsection{Second Order Conductivity for Spontaneous Parametric Down-conversion of X-rays into UV/Visible Radiation}
	Here we derive the expression for the conductivity relevant to the process of SPDC of x-rays into UV/visible radiation.
	In this case, we simplify the expressions by assuming that
	the x-ray wavelengths are far above any electronic transitions and use the dipole approximation for the optical beam (but not for the x rays). We adopt here the notation that is used for SPDC and denote the input x-ray beam, the output x-ray beam, and the optical beam as the pump, signal, and idler respectively.
	
	Since the wavelengths of the x rays are also on the order of the distance between the atomic planes, the reciprocal lattice vector is used to comply with the requirement for momentum conservation (phase matching), 
	given by the equation
	$
	\vb{k}_p+\vb{G}
	=
	\vb{k}_s+\vb{k}_{id}
	$, where
	$\vb{G}$ is the reciprocal lattice vector, and
	$
	\vb{k}_p,
	\vb{k}_s,
	\vb{k}_{id}
	$
	stand for the pump, signal and idler wave vectors respectively.
	Thus the measured intensity is proportional to the Fourier coefficient that corresponds to the selected reciprocal lattice vector.
	The Fourier component of the second order nonlinear conductivity that corresponds to the reciprocal lattice vector $\vb{G}$  of the signal conductivity  
	is 
	\begin{eqnarray}&
	\begin{aligned}
	\sigma_{ijk}(\omega_s;\vb{G})&=
	\frac{e}{i m \omega_p}D_k(-\omega_{id};\vb{G})\delta_{ij}
	\\&
	+\frac{ e^3}{V m^3\omega_{p} \omega_s}
	B_{ijk}(\omega_{id},\vb{k}_{id};\vb{G})	
	,\\& 
	\end{aligned}
	\label{sigma}
	\end{eqnarray}
	where $\omega_s,\omega_{id}$ and  $\omega_{p}$ 
	stand for the signal, idler and pump frequencies
	respectively and $\delta_{ij}$ is the Kronecker delta.
	Here the first term is originated from the the gauge part of the current density operator, and the second term
	is originated from the momentum part of the current density operator, and $D_k(-\omega_{id};\vb{G})$ and $B_{ijk}(\omega_{id},\vb{k}_{id};\vb{G})$ are given by
	\begin{widetext}
		\begin{eqnarray}&
		D_k(-\omega_{id};\vb{G})=
		\left(
		\frac{i\hbar e^2}{mV}
		\right)
		\sum\limits_{n_1\,n_2}
		\bra{W_{n_2}} 
		e^{-i\vb{G}\cdot \vb*{x}} 
		\ket{W_{n_1}}
		\bra{W_{n_1}} 
		\vb{p} \cdot \hat{e}_k
		\ket{W_{n_2}}
		I_{n_2,n_1}(\varepsilon_{id},\vb{k}_{id})
		\label{Dk},
		\\&
		B_{ijk}(\omega_{id},\vb{k}_{id};\vb{G})=
		\sum\limits_{n_1\,n_2}
		\hbar
		(\vb{k}_{id}-\vb{G})
		\cdot
		\bra{W_{n_2}} 
		e^{-i\vb{G}\cdot\vb*{x}}
		\left[
		\hat{e}_i
		(\vb{p}\cdot \hat{e}_j)
		-
		\hat{e}_j
		(\vb{p}\cdot \hat{e}_i)
		\right]
		\ket{W_{n_1}}
		\bra{W_{n_1}} 
		\vb{p} \cdot \hat{e}_k
		\ket{W_{n_2}}I_{n_2,n_1}(
		\varepsilon_{id}
		,\vb{k}_{id})
		,
		\label{Bijk}
		\\&
		I_{n_1,n_2}(\varepsilon_{id},\vb{k}_{id})=
		\frac{V}{(2\pi)^3}
		\int\limits_{B.Z.}d\vb{q}\,
		\frac{
			-\left(f_0(\varepsilon_{n_1}(\vb{q}+\vb{k}_{id}))-f_0(\varepsilon_{n_2}(\vb{q})) \right)
		}{
		\varepsilon_{id}\left[(\varepsilon_{n_1}(\vb{q}+\vb{k}_{id})-\varepsilon_{n_2}(\vb{q}))
		-\varepsilon_{id})+i\hbar\gamma_{n_1\,\vb{q}+\vb{k}_{id},n_2\,\vb{q}}\right]}
	.\label{Ifunction}
	\end{eqnarray}
\end{widetext}
The summation indices $n_1$ and $n_2$ stands for band numbers, 
and the idler energy is given by $\varepsilon_{id}=\hbar\omega_{id}$.
In this work we are interested in referring the nonlinear current density to microscopic and inter unit cell information , thus we use the Wannier basis, which contains this information.
The $\ket{W_{n}}$ is a Wannier function of band $n$.
We relate between the Bloch and Wannier basis by 
\begin{eqnarray}&
\begin{aligned}
\ket{n\,\vb{q}}=
\frac{1}{\sqrt{N}}\sum\limits_{\vb{R}}
e^{i\vb{q}\cdot\vb{R}}T(\vb{R})\ket{W_{n}},
\end{aligned}
\end{eqnarray}
where 
$
T(\vb{R})
=e^
{
	\frac{-i\vb{p}\cdot\vb{R}}{\hbar}
}
$
is the translation operator.

As we show in the appendix, the function $D_k(-\omega_{id};\vb{G})$ is the induced charge density tensor Fourier component.
Each term has a different polarization dependence; while the first term is symmetric with respect to $(i\leftrightarrow j)$, the tensor $B_{ijk}(\omega_{id},\vb{k}_{id};\vb{G})$ is antisymmetric with respect to $(i\leftrightarrow j)$.
This polarization dependence implies that the first term is non-vanishing when the polarization of the signal is parallel to the pump, and the second term is non-vanishing when they are orthogonal to each other. 
The different polarization dependence of the two terms can be used to probe the contributions of the induced charge density, band transitions dependence, and inter-molecular interactions matrix elements.
The functions $B_{ijk}(\omega_{id},\vb{k}_{id};\vb{G})\,$ and $D_k(-\omega_{id};\vb{G}) $ contain the information on the interaction dependence on the idler mode (the long wavelength), the band structure, and inter-molecular interactions matrix elements. 

Equations (\ref{Dk},\ref{Bijk},\ref{Ifunction}) 
have several important and interesting implications. First, by assuming that the Wannier functions are very localized (such that there is no overlap between neighboring functions at different sites), we can separate the contribution of the inter-molecular interactions and the band structure.
The information about the inter-molecular interactions is encoded in the matrix elements and the interaction dependence on the band structure is given by the function $I_{n_1,n_2}(\varepsilon_{id},\vb{k}_{id})$.
It is a measure of the number of band transitions ($n_1 \leftrightarrow n_2$) with separation of $\vb{k}_{id}$ and population difference of $f_0(\varepsilon_{n_1}(\vb{q}+\vb{k}_{id}))-f_0(\varepsilon_{n_2}(\vb{q}))$
that are closely related to the idler energy $\varepsilon_{id}$.
Moreover, it is important to note that for maximally localized Wannier functions (as assumed here) there will only be contributions from interband transitions, when the dipole approximation is assumed with respect to the idler mode.
This is because under this assumption the Wannier functions are real \cite{ri2014proof}. 
\section{Model case of a semiconductor with two bands}

To illustrate the dependence of the conductivity on band structure properties, such as interband transitions and the joint-density of states of interband transitions, it is sufficient to investigate the behavior of the function $I_{n_1,n_2}(\varepsilon_{id},\vb{k}_{id})$.

\begin{figure}[h]
	\includegraphics[scale=0.4]{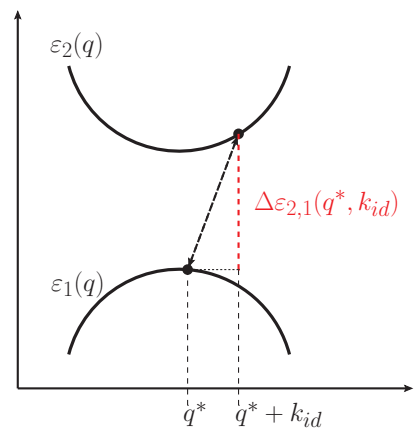}
	\caption{\label{fig:energydifferencebetweenbands} 
		Schematic view of the energy difference between bands ($2\leftrightarrow 1$) at wave vector $q^*$, separated by $k_{id}$.
	}
\end{figure}

To get more insight we consider a simple example of a semiconductor at zero temperature with two energy bands, where the valence band is fully occupied and the conduction band is completely vacant.
Under these assumptions, the difference in the two Fermi-Dirac distributions is proportional to a delta function and therefore, the function $I_{n_1,n_2}(\varepsilon_{id},\vb{k}_{id})$ can be written as a sum of two parts
\begin{eqnarray}&
\begin{aligned}&
I_{n_1,n_2}(\varepsilon_{id},\vb{k}_{id})=\\&\delta_{n_1,2}\delta_{n_2,1}I_+(\varepsilon_{id},\vb{k}_{id})+\delta_{n_1,1}\delta_{n_2,2}I_-(\varepsilon_{id},\vb{k}_{id})
,
\end{aligned}
\end{eqnarray}
where the indices 1 and 2 indicate the valence and conductance band respectively.
By assuming that the damping coefficients are equal to a constant $\gamma$ we obtain
\begin{eqnarray}&
\begin{aligned}&
I_{\pm}(\varepsilon_{id},\vb{k}_{id})=
v
\int d\varepsilon\,
\frac{
	g_{2,1}(\varepsilon,\pm \vb{k}_{id})
}{
\varepsilon_{id}
\left[(\varepsilon\mp\varepsilon_{id})\pm i\hbar\gamma\right]}
,
\end{aligned}
\label{IPM}
\end{eqnarray}
where the integration is performed over states between the two bands separated by a wave vector difference equals to $\vb{k}_{id}$ as illustrated in Fig. \ref{fig:energydifferencebetweenbands}.
\\
$v$ is the volume of the primitive cell, and the weight function in the integral is the joint density of states, which is defined as
$
g_{n_1,n_2}(\varepsilon,\vb{k})=
\frac{1}{(2\pi)^3}
\int\limits_{S(\varepsilon)}
\frac{dS}{
	\abs{
		\grad{}_{\vb{q}}(\Delta \varepsilon_{n_1,n_2}(\vb{q},\vb{k}))
	}
}
$
where 
$
\Delta \varepsilon_{n_1,n_2}(\vb{q},\vb{k})=
\varepsilon_{n_1}(\vb{q}+\vb{k})
-\varepsilon_{n_2}(\vb{q})
$. 
Note that for $\vb{k}=0$ the expression reduces to the standard definition of the joint-density of states \cite{ashcroft1976solid}, and can be considered as such for a wave vector much smaller than a typical size of the first Brillouin zone ($\abs{\vb{k}}\ll \frac{2\pi}{a}$).
From Eq. (\ref{IPM}) and from Fig. \ref{fig:energydifferencebetweenbands},
it is clear that $\varepsilon_{id}$ acts as a resonance in the integral form of the function $I_{+}(\varepsilon_{id},\vb{k}_{id})$, and therefore is more sensitive to the joint-density of states than 
$I_{-}(\varepsilon_{id},\vb{k}_{id})$.
This sensitivity is related to the contribution that arises from the number of energy transitions at the idler energy. 
To illustrate this, we consider the following simple energy dispersion 
\begin{eqnarray}
\begin{aligned}
&
\varepsilon_1 (\vb{k})=V_{ss} \abs{g_1(\vb{k})},
\\&
\varepsilon_2 (\vb{k})=\varepsilon_{\text{gap}}
+V_{ss}(2-\abs{g_1(\vb{k})}),
\end{aligned}
\end{eqnarray}
where $\varepsilon_{\text{gap}}$ is the gap energy, $V_{ss}$ is the width of each band, and $g_1(\vb{k})$ is the band structure form function we used for this example \cite{chadi1975tight}.
The dependence of the functions $I_{\pm}(\varepsilon_{id},\vb{k}_{id}\sim 0)\equiv I_{\pm}(\varepsilon_{id})$ on small idler energies is evaluated numerically under the assumption that the width of the levels are very narrow and therefore the damping coefficient $\gamma$ is neglected.
\begin{figure}[h]
	\includegraphics[keepaspectratio=true,width=1.0\linewidth]{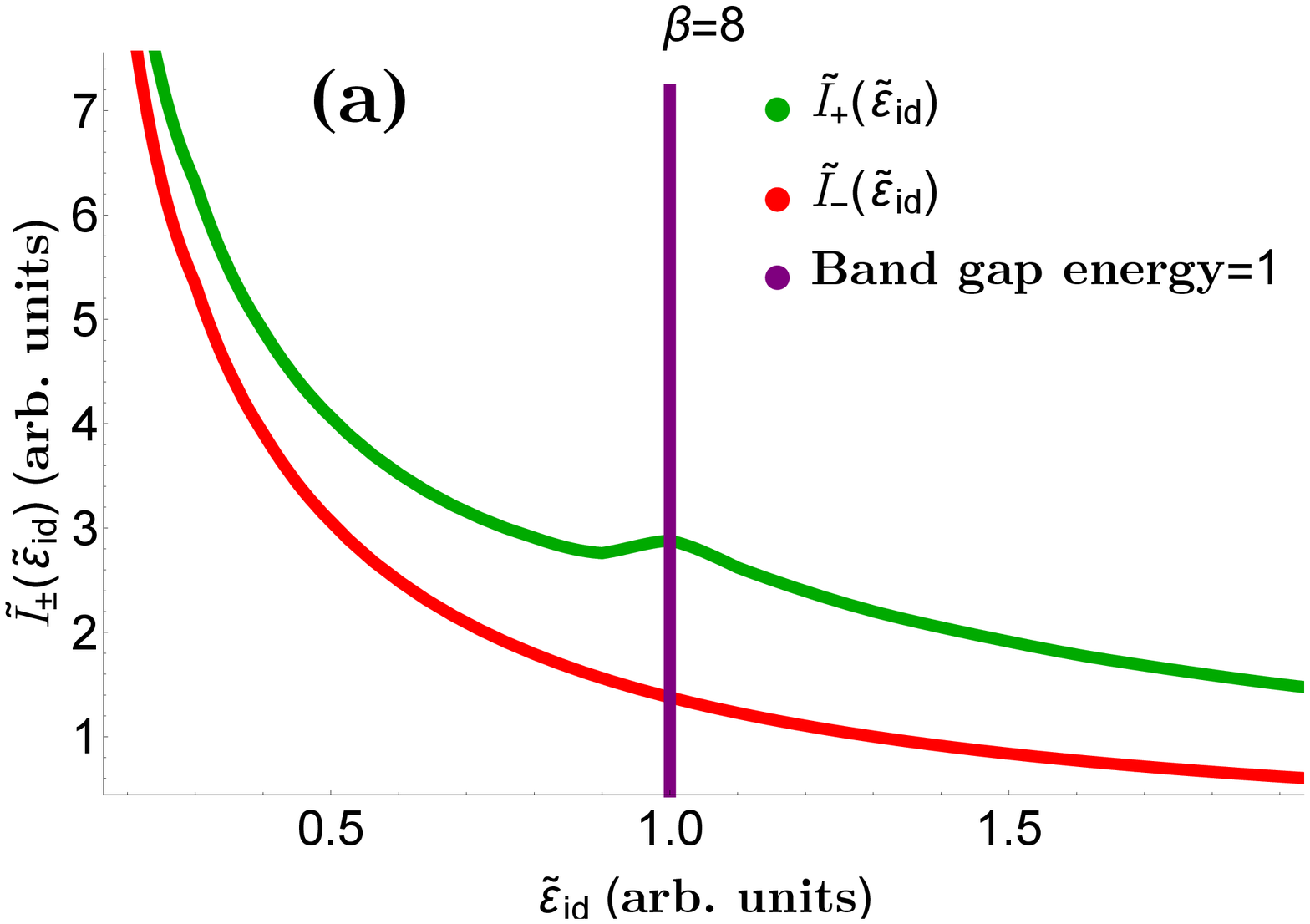}
	\\
	\hfill
	\\
	\includegraphics[keepaspectratio=true,width=1.0\linewidth]{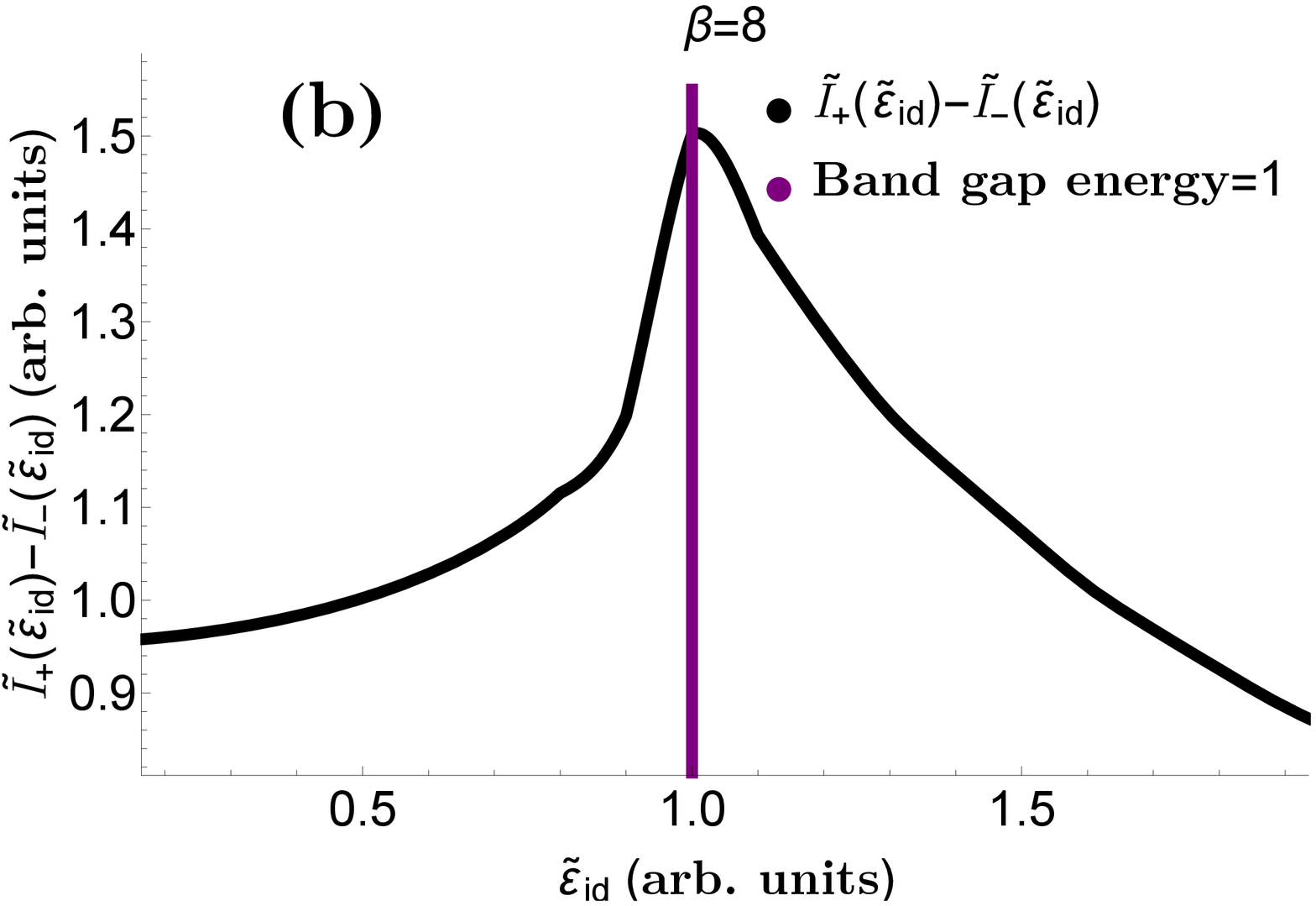}
	\caption{
	Band structure dependence of the nonlinear current density: (a) The term $\tilde{I}_{+}(\tilde{\varepsilon}_{id})$ (green) shows a clear peak at the band gap energy while the term $\tilde{I}_{-}(\tilde{\varepsilon}_{id})$  (red) decreases monotonically with the relative idler energy $\tilde{\varepsilon}_{id}$. (b) The difference between the $\tilde{I}_{+}(\tilde{\varepsilon}_{id})$ and the $\tilde{I}_{-}(\tilde{\varepsilon}_{id})$ term. The vertical purple line indicate band gap energy.  
	}
	\label{fig:bandstructuredependence}	
\end{figure}
Furthermore, in order to see how $I_{\pm}(\varepsilon_{id})$ scale with the band gap energy and to investigate the dependence of the choice of bandwidth $V_{ss}$, we define dimensionless energy parameters 
\begin{eqnarray}
\begin{aligned}
&
\tilde{\varepsilon}_{id}=\frac{\varepsilon_{id}}{
	\varepsilon_{\text{gap}}
}
,\,
\beta=
\frac{2V_{ss}}{\varepsilon_{\text{gap}}}
,\,
\\&
\widetilde{\Delta \varepsilon}(\vb{k})=
\frac{\varepsilon_2 (\vb{k})-\varepsilon_1 (\vb{k})}{\varepsilon_{\text{gap}}}
=
1+\beta
\left(
1-\abs{g_1(\vb{k})}
\right),
\end{aligned}
\end{eqnarray}
where $\tilde{\varepsilon}_{id}$, $\beta$ and $\widetilde{\Delta \varepsilon}(\vb{k})$ are the idler energy, bandwidth energy, and interband transition energy relative to the band gap energy $\varepsilon_{\text{gap}}$ respectively.
With these definitions, the functions $I_{\pm}(\varepsilon_{id})$ can be expressed by dimensionless functions times a scale factor
\begin{eqnarray}
\begin{aligned}
&
I_{\pm}(\varepsilon_{id})=\frac{
	\tilde{I}_{\pm}(\tilde{\varepsilon}_{id})
}
{\varepsilon_{\text{gap}}^2}
,
\\&
\tilde{I}_{\pm}(\tilde{\varepsilon}_{id})=
\frac{v}{(2\pi)^3}
\int\limits_{B.Z.}d\vb{k}\,
\frac{
	1
}{
\tilde{\varepsilon}_{id}
\left(
\widetilde{\Delta \varepsilon }(\vb{k})\mp\tilde{\varepsilon}_{id}
\right)
}
\end{aligned}
\end{eqnarray}
where $\tilde{I}_{\pm}(\tilde{\varepsilon}_{id})$ are dimensionless, and will be used for our analysis.
It is interesting to note that the difference function $I_{+}(\varepsilon_{id})-I_{-}(\varepsilon_{id})$ can be related to the induced charge density when the dipole approximation is assumed with respect to the idler mode. In this case we find that
$
D_k(-\omega_{id};\vb{G})=f_{2,1}^{(k)}(\vb{G}) 
\left(
I_{+}(\varepsilon_{id})-I_{-}(\varepsilon_{id})
\right)
$
where $f_{2,1}^{(k)}(\vb{G})$ encapsulates the inter-molecular interactions with respect to bands 2, 1 and the reciprocal lattice vector $\vb{G}$.  $I_{+}(\varepsilon_{id})-I_{-}(\varepsilon_{id})$ reflects the spectral dependence of the induced charge density on the joint density of states.

We  show the dependence of $\tilde{I}_{\pm}(\tilde{\varepsilon}_{id})$ and of their difference in Fig. \ref{fig:bandstructuredependence}. The prominent peak in $\tilde{I}_{+}(\tilde{\varepsilon}_{id})$ near the band gap energy is due to its strong dependence on the large number of inter-band transitions at the band gap, while $\tilde{I}_{-}(\tilde{\varepsilon}_{id})$ decreases monotonically.
The difference $\tilde{I}_{+}(\tilde{\varepsilon}_{id})-\tilde{I}_{-}(\tilde{\varepsilon}_{id})$ also exhibits a very distinctive peak near the band gap energy.
It is possible to measure this peak if the spectral resolution of the detector is higher than the FWHM of the peak profile.
This enhancement near the band gap energy demonstrates the sensitivity of the nonlinear response and of the induced charge to the joint density of states, and is also in agreement with a recent experiment, showing an enhancement just at the band gap energy \cite{schori2017parametric}.

\begin{figure}[H]
	\includegraphics[keepaspectratio=true,width=1.0\linewidth]{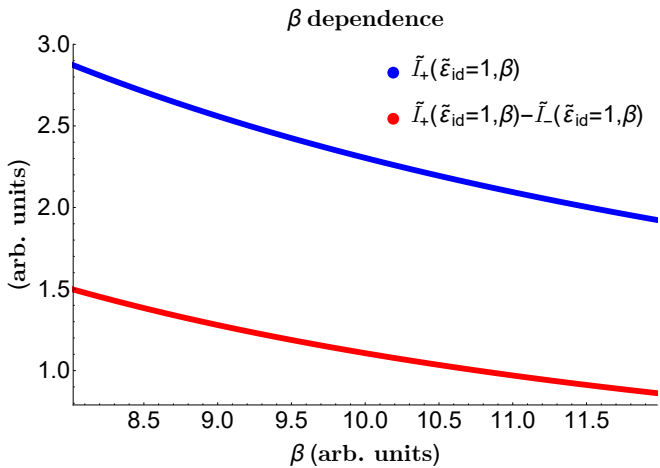}
		\caption{
		$\beta$ dependence of the nonlinear current density:  
		The functions $\tilde{I}_{+}(\tilde{\varepsilon}_{id})$ (blue) and the difference ($\tilde{I}_{+}(\tilde{\varepsilon}_{id})-\tilde{I}_{-}(\tilde{\varepsilon}_{id})$) (red) 
		at the band gap energy ($\tilde{\varepsilon}_{id}=1$). Both decreases with the parameter $\beta$.  
	}
	\label{fig:betadependence}	
\end{figure}

To investigate how the bandwidth effects the model, we evaluated $\tilde{I}_{\pm}(\tilde{\varepsilon}_{id})$ for various values of $\beta=2V_{ss}/\varepsilon_{\text{gap}}$.
The functions $\tilde{I}_{\pm}(\tilde{\varepsilon}_{id})$ display the same behavior for all $\beta$ differing only by a factor.
This difference is displayed in Fig. \ref{fig:betadependence} where the peaks
of the functions decrease with growing values of $\beta$.
This decrease is due to the broadening of the joint density of states, as $\beta$ grows, which leads to a reduction in the density of transitions about the energy gap.
The spectral dependence of the optical response on $\beta$, can be used to probe changes in the band structure due to an external field, and also 
the corresponding changes in $f_{2,1}^{(k)}(\vb{G})$.

In addition to the spectral contribution, we can estimate the contribution of the inter-molecular interactions in the case of two bands by working in a two-dimensional subspace spanned by the Wannier states of each band.
Since the operator $e^{i\vb{G}\cdot\vb*{x}}$ shares the same eigenstates as the position operator, we first find its matrix representation in the eigenbasis of the position operator and then rotate this matrix to the original basis.
Next, we express the matrix elements contained in $D_k(-\omega_{id};\vb{G})$ and $B_{ijk}(\omega_{id},\vb{k}_{id};\vb{G})$ in terms of the position and momentum matrix elements and obtain
\begin{widetext}
	\begin{eqnarray}&
	D_k(-\omega_{id};\vb{G})=-
	\left(
	\frac{i\hbar e^2}{mV}
	\right)
	\pi_k e^{-i\vb{G}\cdot\vb*{a}}\sin(\vb{G}\cdot\vb*{c})
	\left(
	I_+(\varepsilon_{id},\vb{k}_{id})-
	I_-(\varepsilon_{id},\vb{k}_{id})
	\right),
	\\&
	B_{ijk}(\omega_{id},\vb{k}_{id};\vb{G})
	=
	\pi_k e^{-i\vb{G}\cdot\vb*{a}}\cos(\vb{G}\cdot\vb*{c})
	\left[
	\hbar
	(\vb{k}_{id}-\vb{G})
	\cdot
	(
	\hat{e}_i\pi_j
	-
	\hat{e}_j\pi_i
	)
	\right]
	\left(
	I_+(\varepsilon_{id},\vb{k}_{id})
	+
	I_-(\varepsilon_{id},\vb{k}_{id})
	\right),
	\end{eqnarray}
\end{widetext}
where
\begin{eqnarray}
&i\pi_k=\bra{W_{2}} 
\vb{p} \cdot \hat{e}_k
\ket{W_{1}}=
-\bra{W_{1}} 
\vb{p} \cdot \hat{e}_k
\ket{W_{2}},
\\&
\vb*{a}=
\bra{W_{1}} 
\vb*{x} 
\ket{W_{1}}
=
\bra{W_{2}} 
\vb*{x} 
\ket{W_{2}},
\\&
\vb*{c}=
\bra{W_{1}} 
\vb*{x} 
\ket{W_{2}}
=
\bra{W_{2}} 
\vb*{x} 
\ket{W_{1}},
\end{eqnarray}
and the vectors ($\vb*{\pi},\vb*{a},\vb*{c}$) are real.
This form of the matrix elements is attained when assuming that the Wannier functions are real, and that $\comm{x_i}{x_j}=0$.

As an example, we estimate the magnitudes of the nonlinear conductivity for the hydrogen levels 1s and 2p as the two Wannier functions for idler and pump photon energies at 1 eV and 10 keV respectively.
We find that  the first term of the conductivity is on the order of ($\approx 10^{-7} \,\, \frac{\text{A}^3}{\text{W}^2}$). The second term of the conductivity is on the order of ($\approx 10^{-11} \,\,
\frac{\text{A}^3}{\text{W}^2}$).

\section{Summary and Conclusions}

In this work we have shown that the contribution to the nonlinear interaction arises from both band structure properties and atomic-scale interactions. Our results imply that it is possible to extract atomic-scale information on the valence electrons as predicted by previous publications \cite{freund1970optically,freund1972nonlinear,danino1981parametric,tamasaku2011visualizing,glover2012x,rouxel2018x}
but only if the band information can be separated from the Wannier function contribution. For example, when the Wannier functions are localized. Consequently, the spectral dependence of the nonlinearity is essential for the construction of the microscopic information of the electronic states.
Moreover, the population difference between bands or within a band also plays a role in the spectral dependence. When considering more than two bands, there could be an effect of interference between several spectral contributions including interband and intraband transitions, provided that there is a population difference between these transitions.

In contrast to previous publications \cite{glover2012x,borodin2019evidence}, we have found that the polarization of the signal is not only in the direction of the polarization of the pump. 
The two polarization components exhibits different spectral dependencies which can be investigated by measuring each component.

We emphasize that since the nonlinear interaction we discuss is a parametric process in nature (the system does not change its state during the nonlinear interaction) it is inherently ultra-fast. It is therefore very likely that it would be possible to use the nonlinear x-ray and long wavelength interaction for the study of ultra-fast dynamics in solids. Our theory implies that  pump-prob measurements can be  used to study the femtosecond and sub-femotosecond dynamics of interesting processes. Examples include  the variation of the populations between bands, ultra-fast charge transfer between electronic states, optically induced chemical potential variations,  ultra-fast (optically induced) dynamics of band structures, and ultra-fast phase transitions. 

Our formalism can be combined with standard Ab initio methods for further studies of the nonlinear x ray and optical/UV mixing effects in many solid state systems.

We stress that nonlinear interactions can be used to reveal a very broad band spectroscopic information ranging from sub eV to several hundred of eV and structural information of the valence electrons by using a single apparatus. By using SPDC of x-rays to long wavelengths, the energy scan can be done by tuning the angle of the sample and the energy is selected by the detection system.

\begin{acknowledgments}
	The authors thank Emanuele G. Dalla Torre and Efrat Shimshoni for inspiring and useful discussion. This work was supported by the Israel Science Foundation (ISF)(IL), Grant No. 201/17. 
\end{acknowledgments}
\appendix

\section{Induced charge density and the conductivity}
In this section we derive the relation between the induced charge density tensor and the second order conductivity which arises from the so called gauge part of the current density operator.
\\
We start with defining the average charge density to be 
\begin{equation}
\begin{aligned}
\rho_c(\vb*{x},t)&=\tr(
\rho(t)\,
D_{\text{op}}(\vb*{x})
)
\\&=
\sum\limits_{n}\rho_c^{(n)}=
\rho_c^{(0)}(\vb*{x})+\rho_c^{(1)}(\vb*{x},t)+\cdots,
\end{aligned}
\end{equation}
where n signifies the 
order of each term with respect to the field,
such that $\rho_c^{(0)}(\vb*{x})$ is the average charge density in the absence of an external field, and 
$\rho_c^{(1)}(\vb*{x},t)$ is the average induced charge.
The induced charge density can be Fourier expanded in the field modes, namely
\begin{equation}
\begin{aligned}
&
\rho_c^{(1)}(\vb*{x},t)=
\sum\limits_{l}
\rho_c^{(1)}(\vb*{x};\omega_l)
e^{-i\omega_l t}.
\end{aligned}
\end{equation}
The Fourier coefficients of the induced charge density are related to the electric field amplitudes via the charge density tensor
\begin{equation}
\rho_c^{(1)}(\vb*{x};\omega_l)=\sum\limits_{j=1}^{3}
D_j(\vb*{x};\omega_l)
\left(
\vb*{\varepsilon}(\omega_l)
\cdot
\hat{e}_j
\right).
\end{equation}
The charge density tensor has the periodicity of the lattice and therefore can be written as
\begin{equation}
\begin{aligned}
&
D_j(\vb*{x};\omega_l)=
\sum\limits_{\vb{G}}
D_j(\omega_l;\vb{G})
e^{i\vb{G}\cdot\vb*{x}},
\\&
D_j(\omega_l;\vb{G})=
\frac{1}{V}\int\limits_{V}d\vb*{x}\,
D_j(\vb*{x};\omega_l)
e^{-i\vb{G}\cdot\vb*{x}}.
\end{aligned}
\end{equation}
Finally, to relate the gauge part conductivity to the induced charge tensor, we note that
\begin{equation}
\avg{
	\vb*{j}^{(2,\vb{A})}
}(\vb*{x},t)=
\tr(
\rho^{(A)}
\,
\vb*{j}_{\text{op}}^{\vb{A}}
(\vb*{x},t)
)
=\frac{e}{m}\rho_c^{(1)}(\vb*{x},t)\vb{A}(\vb*{x},t).
\end{equation}
Using this relation along with equations (\ref{current-fourierexpansion},\ref{current-conductivity-relation},\ref{conductivity-spatialfouriercomponent}) we find
\begin{equation}
\begin{aligned}&
\sigma_{ijk}^{(2,A)}(\omega_l+\omega_{l'};\vb{G})\\&=
\frac{e}{i m \omega_{l'}}
D_j(\omega_l;\vb{G})
\delta_{ik}
+
\frac{e}{i m \omega_{l}}
D_k(\omega_{l'};\vb{G})
\delta_{ij}.
\end{aligned}
\end{equation}


\begin{references} 
	\bibitem{danino1981parametric}
	H. Danino and I. Freund, Physical Review Letters \textbf{46},
	1127 (1981).
	\bibitem{tamasaku2009determining}
	K. Tamasaku, K. Sawada, and T. Ishikawa, Physical
	review letters \textbf{103}, 254801 (2009).
	\bibitem{tamasaku2011visualizing}
	K. Tamasaku, K. Sawada, E. Nishibori, and T. Ishikawa,
	Nature Physics \textbf{7}, 705 (2011).
	\bibitem{glover2012x}
	T. Glover, D. Fritz, M. Cammarata, T. Allison, S. Coh,
	J. Feldkamp, H. Lemke, D. Zhu, Y. Feng, R. Coffee, et al.,
	Nature \textbf{488}, 603 (2012).
	\bibitem{barbiellini2015explaining}
	B. Barbiellini, Y. Joly, and K. Tamasaku, Physical Review
	B \textbf{92}, 155119 (2015).
	\bibitem{schori2017parametric}
	A. Schori, C. B{\"o}mer, D. Borodin, S. P. Collins, B. Detlefs,
	M. MorettiSala, S. Yudovich, and S. Shwartz, Physical review
	letters \textbf{119}, 253902 (2017).
	\bibitem{rouxel2018x}
	J. R. Rouxel, M. Kowalewski, K. Bennett, and
	S.Mukamel, Physical Review Letters \textbf{120}, 243902 (2018).
\bibitem{borodin2019evidence}
D. Borodin, A. Schori, J.-P. Rueff, J. M. Ablett, and
S. Shwartz, Physical review letters \textbf{122}, 023902 (2019).
\bibitem{sofer2019observation}
S. Sofer, O. Sefi, E. Strizhevsky, S. Collins, B. Detlefs,
C. J. Sahle, and S. Shwartz, arXiv preprint
arXiv:1904.13146 (2019).
	\bibitem{tamasaku2007idler}
	K. Tamasaku and T. Ishikawa, Acta Crystallographica
	Section A: Foundations of Crystallography \textbf{63}, 437
	(2007).
\bibitem{borodin2017high}
D. Borodin, S. Levy, and S. Shwartz, Applied Physics
Letters \textbf{110}, 131101 (2017).
    \bibitem{sinha2001theory} 
    S. K. Sinha, Journal of Physics: Condensed Matter \textbf{13},
    7511 (2001).
   \bibitem{wang2012lindhard}
Y. J. Wang, B. Barbiellini, H. Lin, T. Das, S. Basak,
P. E. Mijnarends, S. Kaprzyk, R. S. Markiewicz, and A. Bansil,
Physical Review B \textbf{85}, 224529 (2012).
	\bibitem{baron2015introduction}
	A. Q. Baron, arXiv preprint arXiv:1504.01098 (2015).
	[10] I. Freund and B. Levine, Physical Review Letters \textbf{25},
	1241 (1970).
	\bibitem{freund1970optically}
	I. Freund and B. Levine, Physical Review Letters \textbf{25},
	1241 (1970).
	\bibitem{freund1971resonant}
	I. Freund and B. Levine, Optics Communications \textbf{3}, 101
	(1971).
	\bibitem{freund1972nonlinear}
	I. Freund, Chemical Physics Letters \textbf{12}, 583 (1972).
	\bibitem{dorfman2012photon}
	K. E. Dorfman and S. Mukamel, Physical Review A \textbf{86},
	023805 (2012).
	\bibitem{freund1968nonlinear}
	I. Freund, Physical Review Letters \textbf{21}, 1404 (1968).
	\bibitem{jha1968nonlinear}
	S. Jha and C. Warke, Il Nuovo Cimento B (1965-1970)
	\textbf{53}, 120 (1968).
	\bibitem{freund1969parametric}
	I. Freund and B. Levine, Physical Review Letters \textbf{23}, 854
	(1969).
	\bibitem{jha1972nonlinear1}
	S. S. Jha and J. W. Woo, Physical Review B \textbf{5}, 4210
	(1972).
	\bibitem{jha1972nonlinear2}
	S. Jha and J. Woo, Il Nuovo Cimento B (1971-1996) \textbf{10},
	229 (1972).
	\bibitem{popova2018theory}
	D. Popova-Gorelova, D. A. Reis, and R. Santra, Physical
	Review B \textbf{98}, 224302 (2018).
	\bibitem{bloembergen1964quantum}
	N. Bloembergen and Y. Shen, Physical Review \textbf{133}, A37
	(1964).
	\bibitem{callaway2013quantum}
	J. Callaway, \textit{Quantum theory of the solid state} (Academic
	Press, 2013).	
	\bibitem{ri2014proof}
	S. Ri and S. Ri, arXiv preprint arXiv:1407.6824 (2014).
	\bibitem{ashcroft1976solid}
	N. W. Ashcroft and N. D. Mermin, \textit{Solid state physics}
	(Cengage Learning, 1976).
	\bibitem{chadi1975tight}
	D. Chadi and M. L. Cohen, physica status solidi (b) \textbf{68},
	405 (1975).
\end{references}
%

\end{document}